# Long-Range Superconducting Proximity Effect in Nickel Nanowires

Jue Jiang[1], Weiwei Zhao[1], Fei Wang[1], Renzhong Du[1], Ludi Miao[1], Ke Wang[2], Qi Li[1], Cui-Zu Chang[1], and Moses H.W. Chan[1]

[1]Department of Physics, The Pennsylvania State University, University Park, PA 16802

[2]Materials Research Institute, The Pennsylvania State University, University Park, PA 16802

Corresponding author: mhc2@psu.edu (M. H. W. C.)

**Abstract: When a ferromagnet is placed in contact with a superconductor, owing to incompatible spin order, the Cooper pairs from the superconductor cannot survive more than one or two nanometers inside the ferromagnet. This is confirmed in the measurements of ferromagnetic nickel (Ni) nanowires contacted by superconducting niobium (Nb) leads. However, when a 3 nm thick copper oxide (CuO) buffer layer made by exposing an evaporated or a sputtered 3 nm Cu film to air, is inserted between the Nb electrodes and the Ni wire, the spatial extent of the superconducting proximity range is dramatically increased from 2 to a few tens of nanometers. Scanning transmission electron microscope study confirms the formation of a 3 nm thick CuO layer when an evaporated Cu film is exposed to air. Magnetization measurements of such a 3 nm CuO film on a $SiO_2$/Si substrate and on Nb/$SiO_2$/Si show clear evidence of ferromagnetism. One way to understand the long-range proximity effect in the Ni nanowire is that the CuO buffer layer with ferromagnetism facilitates the conversion of singlet superconductivity in Nb into triplet supercurrent along the Ni nanowires.**

**Main text**: The leakage of Cooper pairs extends superconducting behavior into a normal metal when it is placed in direct contact with a superconductor. The spatial range of such a proximity effect can be as long as 1 µm[1,2]. However, when the normal metal is replaced by a ferromagnet, the proximity-induced superconductivity is expected to decay rapidly inside the ferromagnet and vanish within one or two nanometers due to the incompatible nature of superconductivity and ferromagnetic order[2]. This expectation was confirmed in macroscopic (Fe, Ni)-In junctions[3] and submicron Ni-Al structures[4] where the spatial range of the proximity effect is found to be ~ 1 nm. On the other hand, a surprisingly long-ranged proximity effect was found in a number of experiments on mesoscopic superconductor-ferromagnet (S-F) hybrid structures[5-12]. Supercurrent was detected in a half-metallic ferromagnet $CrO_2$ thin film sandwiched between two singlet superconducting electrodes separated by 1 µm[10,13]. To account for the unusually long-ranged proximity effect, the induced superconductivity in the $CrO_2$ film was attributed to spin-triplet pairing instead of the usual singlet state. Josephson junctions with tailored magnetic profiles were fabricated to test this spin-triplet generation mechanism, where the ferromagnetic Co layer is sandwiched between two conical magnetic Holmium (Ho) layers of 4.5 and 10 nm thick[7], whose magnetic moments follow a helical pattern along the c-axis at low temperatures. It displayed a relatively constant voltage, which is the critical current multiply normal-state resistance ($I_cR_N$), as a function of the thickness when the ferromagnetic Co layer is up to 16 nm. In the absence of the Ho layers, $I_cR_N$ would decrease exponentially. Similar results [8,9] have been found when a weak ferromagnetic layer PdNi (2.8 nm) or CuNi (1.4 nm) were inserted as buffer layers between Nb and Co/Ru/Co. In these systems, $I_cR_N$ remains constant when the Co layer is increased to 28 nm, indicating the robustness of spin-triplet superconductivity against the presence of ferromagnet.

Similar spin-triplet-induced long-range superconductivity is recently observed in the Josephson junction consists of a newly discovered triangular magnet, $K_{1-x}V_3Sb_5$[14].

A long-range proximity effect was also reported in ordinary hard ferromagnetic Co and Ni nanowires (NW) contacted with superconducting tungsten (W) electrodes[15]. The nanowires were grown electrochemically inside a porous membrane and "harvested" individually for measurements. The electrodes, containing approximately 40% atomic carbon and 20% atomic gallium were deposited onto the NWs by focused ion beam (FIB) technique and have a superconducting transition temperature $T_c$ of ~4.5 K, well above that of pure W at $T$ ~12 mK. The key findings, measured with a 4-probe configuration, are that a Co NW of 40 nm diameter was driven completely superconducting with zero resistance when sandwiched between two superconducting W voltage electrodes separated by 600 nm. For longer Co NWs of 40 and 80 nm diameter and 1.5 µm length and Ni NW of 60 nm diameter and 3 µm long, the residual resistance found after the superconducting drop in the low-temperature limit were 11, 50, and 52 percent respectively of the normal state resistance. The long-range proximity effect is also seen in a configuration where a superconducting W strip is placed in contact with the Co NW but is not part of the measurement circuit. The deposition of the W electrodes onto the NWs by the FIB deposition process involves the bombardment of high-energy ions of the NW. This process very likely produces defects and inhomogeneous magnetic moments in W/Co and W/Ni contact regions. It was proposed that[10,16-18] the conversion from singlet to triplet superconductivity requires inhomogeneous magnetic moments. It is reasonable to interpret the long-range proximity effect in Co and Ni NW to be a consequence of triplet superconductivity induced by the inhomogeneous magnetic moment at the W/Co and W/Ni contact regions. The results of Ref. [15] were later replicated by another group[19]. The major drawback of these two experiments is that the deposition of the W electrodes by FIB is not a well-controlled process. The experimental studies[5-15,19] cited above generated considerable excitement in the condensed matter community since spin-triplet superconductivity, including that generated by proximity effect is thought to host topological excitations that may be utilized for quantum computations[17].

Here we report an experiment that shows the spatial extent of the proximity effect along a Ni NW is dramatically lengthened by nearly two orders of magnitude upon the insertion of a 3 nm thick naturally oxidized thin Cu buffer layer between the superconducting Nb electrodes and the NW. In contrast, the insertion of an Au buffer layer or 3 nm Cu buffer layer prevented from oxidation gives rise to a much smaller superconducting proximity range. Additionally, the insertion of 10 nm thick Cu with and without natural oxidation shows no enhancement in the proximity distance. Magnetometry measurements on these different Cu films indicate that only the 3 nm thick oxidized Cu buffer layer exhibits ferromagnetic property.

In our experiment, samples were fabricated by e-beam lithography followed by physical vapor deposition (PVD). Materials with different functionality were deposited separately to assure comparatively clean interfaces with minimal intermixing. The superconducting Nb electrodes (500 nm (wide)×40 nm (thick)) in all samples were made by DC magnetron sputtering on the Si/SiO$_2$ substrates. Consistent superconducting transition temperature ($T_c$) is found near 8 K. The background vacuum of the sputtering process is ~5×10$^{-7}$ mbar, and Argon pressure is 4×10$^{-3}$ mbar for exciting the plasma. The deposition of the ferromagnetic Ni or Cu buffer layer is carried out either by sputtering or thermal evaporation with a background vacuum of ~ 1×10$^{-6}$ mbar. These PVD-made NWs typically have a polycrystalline structure[20]. The schematic and optical images for a typical transport measurement circuit are shown in the inset of Fig. 1a. Low-temperature transport measurements are carried out in the Physical Property Measurement System (PPMS) and its auxiliary dilution fridge inserts from Quantum Design.

Figure 1a shows 4-probe magneto-transport measurements of *Sample 1*, a Ni NW (400 nm (wide) × 40 nm (thick)) deposited by thermal evaporation, followed by exposure to air during a second nano-fabrication step prior to the sputtering of superconducting Nb electrodes (500 nm (wide) × 40 nm (thick)) on top of the oxidized Ni NW. The observed small drops in resistance at low temperature (Fig. 1a) and low magnetic field (Fig. 1b) are signatures of superconductivity in the magnetic Ni nanowire in contact with the superconducting Nb voltage leads. Since the total length of the Ni NW between the voltage leads is 500 nm and the total drop in resistance seen under zero magnetic field is ~ 0.9% at $T$ ~0.5 K, the spatial range of the proximity effect in the Ni NW is estimated to be ~ 2 nm, in a good agreement with theoretical expectations[1,2]. Figure 1b also shows, as expected, that the proximity-induced resistance-drop decreases with the application of an out-of-plane external magnetic field and elevated temperature. $R$-$\mu_0 H$ relations at $T$ ~8 K indicate an anisotropic magnetoresistance (AMR) on top of the superconducting resistance-drop, giving rise to a "step" feature in magnetoresistance upon reversing the direction of the magnetic field. This resistance difference due to AMR does not exceed ~0.4% of average resistance. The inset in Fig. 1b shows the 2-terminal resistance measurement under zero magnetic field. The large 2-terminal resistance (3.4 kΩ above 8 K and nearly 2 kΩ below 4 K) and the substantial scatters in the four terminal resistance value indicates there is an insulating Ni oxide layer between the Nb electrodes and the Ni NW. The drop in 2-terminal resistance near 8 K pinpoints the superconducting transition of at least one of the Nb electrodes at this temperature. The additional resistance drop seen near 4 K may locate the $T_c$ of other Nb electrodes. In addition, the inset of Fig. 1b shows an upturn in the two terminal resistance value below 2K, this upturn is very likely the signature of Mott insulator behavior of the Ni oxide layer.

In Figs. 1c and 1d, we present transport results of *Sample 2* which shows a very small 2-terminal resistance (305 Ω above and 20 Ω below 8K). For this sample, a 3 nm thick Au layer was evaporated onto NW prior to sputtering the Nb electrodes. Both *R-T* and *R*-$\mu_0 H$ scans show "critical peaks" near the superconducting transition of the Nb electrode. These peaks have been reported in prior studies[13-15,21,22] but the physical origin of these peaks is not yet clear. Below the superconducting transition of the Nb leads, the resistance of Ni NW drops rapidly by ~4% and flattened out quickly with decreasing *T* and external magnetic field. The 4% drop translates to a superconducting proximity range of ~10 nm, indicating that the insertion of the Au buffer layer while improves very significantly the contact transparency between Nb leads and the Ni NW the enhancement of the spatial range of the proximity effect is modest. In another sample, we evaporated a 6 nm Au film over the entire length of the Ni NW before sputtering Nb leads (*Sample 8*, Supplementary S2) [23]. This sample exhibits a nearly identical proximity range of 10 nm. This result suggests the ferromagnetism of the Ni NW prevented the thin Au layer along its entire length from being driven to be superconducting by the Nb leads[24,25].

Figure 2 shows magneto-transport measurements of two different Ni NW samples contacted with Nb leads through a thin CuO buffer layer. In *Sample 3*, Ni NW was thermally evaporated onto the Si/SiO$_2$ substrate, it was then taken out from the vacuum for the second lithography step to prepare for the 3 nm Cu film evaporation. The sample was exposed to air again to oxidize the Cu film and finally placed in the sputtering chamber where Nb leads were attached. For *Samples 4*, Nb electrodes and the 3 nm Cu film were sputtered sequentially without breaking vacuum, then the sample was taken out from the vacuum and went through the second lithography step for Ni

NW sputtering. Since the Cu films in both samples were exposed to air for a few hours, there is inevitably a CuO layer between the Nb electrodes and the Ni NWs in both Samples. Our scanning transmission electron microscopy and energy-dispersive X-ray spectroscopy studies, to be presented below indicate a Cu layer of 3 nm thick similarly exposed to air is likely to be oxidized. Since the Nb and Cu films of *Sample 4* were sputtered sequentially without breaking the vacuum, there may be an oxide-free Nb/Cu interface between the Nb electrode and the CuO buffer layer in *Sample 4*.

Since the voltage leads of *Sample 3* are separated by a distance $L$ of 500 nm the resistance-drop of 30% between 3 K to 0.2 K, indicating a superconducting proximity spatial range of 65 nm. This means the insertion of a 3 nm CuO layer between the Ni NW and the Nb electrode increases the spatial proximity effect by a factor of 30. The magnitude of the resistance- drop and the "critical" field value of the phenomenon shows the expected temperature dependence (Fig. 2b). The 2-terminal results in the inset of Fig. 2b show a drop near 8 K due to the superconducting transition of Nb electrodes. 2-terminal resistance values of *Samples 3* and *4*, near 30 and 45 k$\Omega$ respectively, like that in *Sample 1*, are also much higher than that of the 4-terminal values. For *Samples 3* and *4* the two terminal values, which include the resistances of the 3 nm insulating CuO buffer layers sandwiched between the Ni NW and the two Nb electrodes, show prominent upturn below 3 K (*Sample 3*) and 2 K (*Sample 4*). These upturns as noted above are low temperature signatures of Mott insulator behavior of the CuO buffer layers. The 2-terminal resistance values of these two samples are much higher than that of *Sample 1* because the thickness of the 3 nm CuO buffer layers is much thicker than that of the nickel oxide layer between the Nb electrodes and the Ni NW in *Sample 1*.

Figures 2c and 2d show magneto-transport results of *Sample 4*. In contrast to *Samples 1* and *3*, the resistance of *Sample 4* begins to drop promptly at $T \sim 8$ K, i.e., $T_c$ of Nb, instead of at a lower temperature. The magnitude of the resistance drop of *Sample 4* indicates a proximity range of 136 nm. In addition, a much larger current than that used in Sample 3 is required to quench completely the proximity-induced superconductivity (see Supplementary S1) [23]. The longer superconducting proximity range in *Sample 4* is likely due to the aforementioned oxide-free Nb/Cu interface that is more favorable for Andreev reflection. "Critical peaks" near $T_c$ and $H_c$ as well as a repeatable quasiperiodic oscillation superimposed on background magnetoresistance are observed in *Sample 4*. These oscillations are likely due to the crossing of vortices in the proximity-induced superconducting region[26].

The long superconducting proximity range shows, interestingly, a very significant decrease when we prevent air exposure or increase the thickness of the Cu buffer layer. In *Sample 5,* we first sputtered Ni NW, then broke the vacuum for the second lithography step to prepare for sequential sputtering of Cu (3 nm) and Nb film in the same vacuum chamber without further breaking the vacuum. Such a process eliminated the exposure of the Cu layer to the ambient atmosphere. We saw a superconducting proximity range of only 13 nm at 2 K (See Supplementary S2)[23]. We repeated the process used for *Sample 5* to fabricate *Sample 6* but increased the Cu thickness to 10 nm, resulting in a proximity range of only 2 nm. For *Sample 7*, we purposely exposed the 10 nm Cu buffer layer to air prior to the deposition of Nb, and still found a proximity range of less than 1 nm. We have summarized the preparation procedures of all the NW samples presented in this *Letter* in Table 1 for easy reference.

The results we have presented indicate the long proximity range is related to the oxidation of the thin, ~ 3 nm thick Cu buffer layer in the fabrication process. In contrast to Cu oxide, the Ni oxide layer found at the Ni/Nb interface in *Sample 1* does not lengthen the proximity range. Interestingly, when the Cu buffer layer between Nb and Ni is increased to 10 nm, either allowed to be oxidized, as in the case for *Sample 7,* or prevented from oxidation as in the case for *Sample 6*, the proximity remains to be short-range at 2 nm.

The existence of the CuO layer is clearly revealed by high-angle annular dark-field scanning transmission electron microscopy (HAADF STEM) and energy-dispersive X-ray spectroscopy (EDS) studies. The specimen is made by evaporating a 12 nm Cu film onto Si substrate, followed by air exposure for more than 8 hours. It was then transferred to the sputtering chamber for Nb deposition. Figures 3a and 3b show a Nb layer, along with the Cu buffer layer highlighted by the yellow dashed box. Since heavier elements are brighter in the HAADF STEM images, the "dark" layer below the Nb layer represents the oxidized Cu. This Cu oxide layer is more clearly revealed by EDS elemental mapping in Fig. 3c, where oxygen is found extending between 3 and 4 nm into the 12 nm Cu film. Therefore, the 3 nm Cu buffer layers in our NW samples (namely *Samples 3* and *4*) that reveal a long-range proximity effects are very likely to be fully oxidized and the 10 nm Cu film in *Sample 7* is only partially oxidized during air exposure.

To correlate the transport results with magnetic properties of the Cu buffer layer, we grew naturally oxidized 3 nm Cu film, 3 nm Cu film prevented from oxidation, and naturally oxidized 10 nm Cu film onto 4 mm × 4 mm $SiO_2$ (500 nm)/Si (500 µm) substrates. The protocols of how these film samples are made are shown in Table 2. A superconducting quantum interference device (SQUID) based Magnetic Property Measurement System (MPMS, Quantum Design) was used to measure the magnetic properties of these samples with the 4 mm × 4 mm substrates intact and oriented perpendicular to the magnetic field. We also carried out measurements on identical 'pristine' substrates and on substrates sputtered with Nb film. The $SiO_2$/Si substrate, labeled as *Sample MH1*, exhibits a signature diamagnetic property with a negative linear magnetic field dependence of magnetization (*M*) on *H* (Fig. 4a). We followed the same procedure used in growing the Cu buffer layer in *Sample 3* and *Sample 4* by depositing a 3 nm Cu layer on the $SiO_2$/Si substrate, followed by air exposure for more than 8 hours. The *M-H* plots of this sample (*Sample MH2*) at 10 and 50 K are shown in Fig. 4b. For $|H| > 0.5$ T, the plots of *M* show negative linear dependence on *H*, or diamagnetic behavior, just like that found for *Sample MH1*, the $SiO_2$/Si substrate. For $|H| < 0.2$ T, the plots show inverted sigmoid deviations with hysteresis in addition to the diamagnetic behavior. To reveal the *M-H* response of *Sample MH2* in the low field region, we isolate the diamagnetic contribution of the samples. The diamagnetic contribution is obtained by extrapolating the two branches of the linear and parallel *M* vs. *H* traces, for $|H| > 0.5$ T and 'translate' them along the *x* or *H* axis, without changing the slope to go through the origin. Figure 4c shows the *M* vs. *H* response of *Sample MH2* (Fig. 4b) with the diamagnetic contribution subtracted. The figure shows standard hysteretic ferromagnetic behavior with coercive field of 0.2 T. The magnetization also shows an increase with decreasing temperature, and the saturated magnetization at 10K is found to be $4.8\times 10^{-6}$ emu. Since the ferromagnetism originated from the CuO layer on the $SiO_2$/Si substrate, it would be sensible to scale the magnetization by the surface area of the substrate (4 mm × 4 mm) instead of the volume of the CuO film. The per unit area of saturation magnetization ($M_s$) of the CuO film is found to be $3\times10^{-5}$ emu/cm$^2$. The ferromagnetism of the 3 nm CuO film resembles recent findings on thin $VSe_2$ films. Bulk $VSe_2$ is non-ferromagnetic bulk but monolayer $VSe_2$ is weakly ferromagnetic[27,28]. The measured $M_s$ (by area) is found to be $2.4\times10^{-4}$ emu/cm$^2$ [27], or one order of magnitude larger than the 3 nm CuO film on *Sample MH2*.

We note that the *M-H* plots of Fig. 4c show only the ferromagnetic behavior without the paramagnetic contribution expected for a magnetic material. The reason for this is that the paramagnetic term scales linearly with *H*, just like the diamagnetic contribution from the $SiO_2$/Si substrate, with an opposite sign. Since the CuO film is 3 nm thick or $1.5 \times 10^5$ times thinner than the 500 µm $SiO_2$/Si substrate, the expected paramagnetic contribution is completely buried by and has been 'subtracted' away with the diamagnetic substrate background.

In order to ascertain the finding of ferromagnetic behavior of 3 nm CuO film, we made magnetic measurements on two additional samples prepared with the same procedures as *Sample MH2* (See Supplementary S3)[23]. All three samples, *Samples MH2*, *MH2a* and *MH2b* show similar ferromagnetic behavior. The saturation magnetizations of these samples are respectively $3.0 \times 10^{-5}$, $2.5 \times 10^{-5}$ and $1.25 \times 10^{-5}$ emu/cm$^2$. The variations reflect the uncontrolled oxidation process of the Cu films. We have also made measurements on a 3 nm oxidized Cu layer deposited on top of the 40 nm Nb layer (*Sample MHS2* in Supplementary S3)[23]. Similar ferromagnetic behavior is found indicating the ferromagnetism is not affected by the Nb layer. On the other hand, Figure 4d shows that there is no evidence of any ferromagnetic response in *Sample MH3*, a thicker 10 nm Cu layer that is partially oxidized as demonstrated by the STEM image. Similarly, a 3 nm Cu film sputtered onto Nb layers without breaking vacuum (*Sample MHS3* in Supplementary S3)[23], also shows no ferromagnetic response.

The magnetization results shown in Fig. 4 and Fig. S5 show direct correlation between long range proximity effect (*Samples 3* and *4* in Fig. 2a and 2c) and the ferromagnetism of the 3 nm CuO buffer layer (*Sample MH2* in Fig. 4c). For thicker, not oxidized or partially oxidized Cu buffer layer, the absence of ferromagnetism (*Sample MH3* in Fig. 4d, and *Sample MHS3* in Fig. S5c) is correlated with the absence of long range proximity effect (*Samples 6 and 7*, in Fig. S3a and Fig.

S3c). As noted above, in *Sample 5*, the oxidation of the 3 nm Cu film was prevented because the sputtering of the Cu film and the Nb leads occurred sequentially with the same mask without breaking the vacuum. As expected, in a 3 nm thick Cu film prevented from oxidation (*Sample MHS3*, Fig. S5c) no evidence of ferromagnetism was found. However, a modest (13 nm) proximity range was found in *Sample 5*. We think this modest proximity range may be related to the fact that the edges of the Cu buffer layer in *Sample 5* not covered by the Nb electrodes are oxidized and become ferromagnetic. The Cu film near the edges of *Sample MH3* used for magnetization measurements is also likely oxidized and become ferromagnetic. However, the fraction of the Cu film at the edges to that in the interior of the 4 mm × 4 mm sample is miniscule and cannot be picked up by magnetization measurement. The edge to interior fraction of the Cu layer in *Sample 5* can be many orders larger than that of *Sample MH3*. The one-to-one correlation between the ferromagnetism of the fully oxidized Cu buffer layer and the long spatial superconducting proximity effect in Ni NW demonstrate a causal relationship between these two phenomena.

This ferromagnetic behavior in the naturally oxidized Cu layer of different thickness agrees with the ferromagnetic response reported for nanoparticles or thin-film CuO [29-34]. In CuO nanoparticles, weak ferromagnetism is attributed to uncompensated surface spins. Ref. [33] reports the size-dependent magnetic property of CuO nanoparticles. Specifically, it reports that anti-ferromagnetism is dominant when the diameter ($d$) is larger than 10 nm, but weak ferromagnetism emerges when $d$ is smaller than 10 nm. Ref. [31,32] shows that the oxygen vacancies is a source of the uncompensated spins. Based on Néel's model [35,36], the magnetic moment of a nano-structure depends on the inequality of antiferromagnet's two anti-parallel sublattices, which are influenced by morphology, crystal structure, and size. When the alternating sublattice "neutral" planes have incomplete top and bottom surfaces, the magnetic moment will be inversely proportional to the dimension of the nanostructure. The ferromagnetism in our 3 nm thick oxidized Cu film appears to share the same physical origin in these CuO nanostructures. As noted above, the ferromagnetism of the 3 nm CuO film resembles that of thin $VSe_2$ films.

In summary we have presented results that show the thin ferromagnetic oxidized Cu buffer layer is responsible for the possible spin-triplet superconductivity with long proximity spatial range along the length of ferromagnetic Ni NW. Since the deposition and natural oxidation in the ambient atmosphere of a thin, ~3 nm Cu layer between superconducting electrodes and a ferromagnetic NW is a simple and easily reproducible process, the results presented here open an easily accessible procedure for generating spin-triplet superconductivity for systematic in-depth studies.

**Acknowledgments:** The authors would like to thank C. X. Liu for helpful discussions and Z. Q. Wang, J. J. Wang, S. Kempinger, and M. Kayyalha for the assistance in our experiments. J. J. and M. H. W. C. acknowledge the support from NSF Penn State MRSEC Grant DMR-1420620 and NSF grant DMR 1707340. Q.L. acknowledges the support from DOE (DE-FG02-08ER46531) for superconducting film fabrication and NSF DMR- 1905833. F. W. and C.-Z. C. acknowledge the support from the NSF-CAREER award (DMR-1847811) and the Gordon and Betty Moore Foundation's EPiQS Initiative (Grant GBMF9063 to C. -Z. C.).

**Data Availability**：The data that support the findings of this study are available from the corresponding author upon reasonable request.

**Figures and Tables**

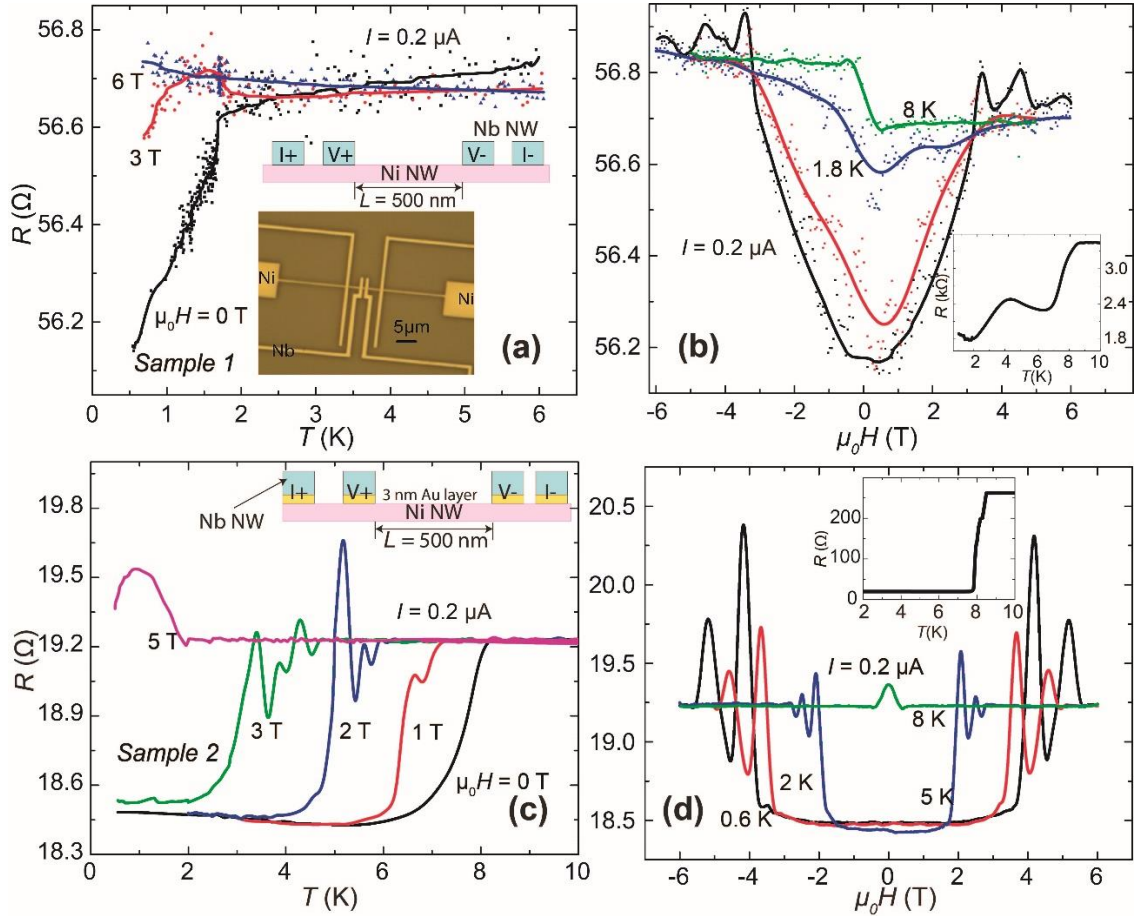

**Figure 1 | $R$-$T$ and $R$-$\mu_0 H$ curves of Ni nanowires contacted by superconducting Nb electrodes without or with an Au buffer layer.** (a) In *Sample 1*, Ni NW (pink) was thermally evaporated on the $SiO_2$/Si substrate and then exposed to air and moved to a different chamber for sputtering of tNb electrodes (blue). At zero magnetic field, the resistance $R$ starts to drop at $T$ ~1.5 K. Since the drop at $T$ ~0.5 K is ~0.8 % of the resistance of the normal state, the superconducting proximity range is ~ 2 nm. Inset: a schematic drawing and an optical image of *Sample 1*. (b) $R$-$\mu_0 H$ curves of *Sample 1* under different temperatures. Inset: the 2-terminal resistance shows two drops at $T$ ~8 K and $T$ ~4 K, possibly indicating different $T_c$ of the two Nb electrodes. An upturn due to the Mott insulator behavior of Ni oxide is found below 2K. In (a) and (b), the dots are the raw data and the lines are guides to the eye. (c) In *Sample 2*, Ni NW (pink) was thermally evaporated, then it was exposed to air, followed by the second-step lithography and Au (yellow) evaporation. It was taken

out from the vacuum and placed in a sputtering chamber for Nb (blue) deposition. At zero magnetic field, the resistance $R$ drop of ~4% corresponds to a superconducting proximity range of ~10 nm, indicating a slight enhancement in the spatial range of supercurrent in Ni NW with an Au buffer layer. (d) $R$-$\mu_0 H$ curves of *Sample 2* under different temperatures. Inset: the 2-terminal resistance of *Sample 2*. Both $R$-$T$ and $R$-$\mu_0 H$ relations show several "critical peaks" near the superconducting transition regimes, consistent with our prior report [15]. The fabrication details of *Sample 1* and *Sample* 2 are listed in Table 1. Ni NW used here is 400 nm wide and 40 nm thick, while the Nb electrodes are 500 nm wide and 40 nm thick.

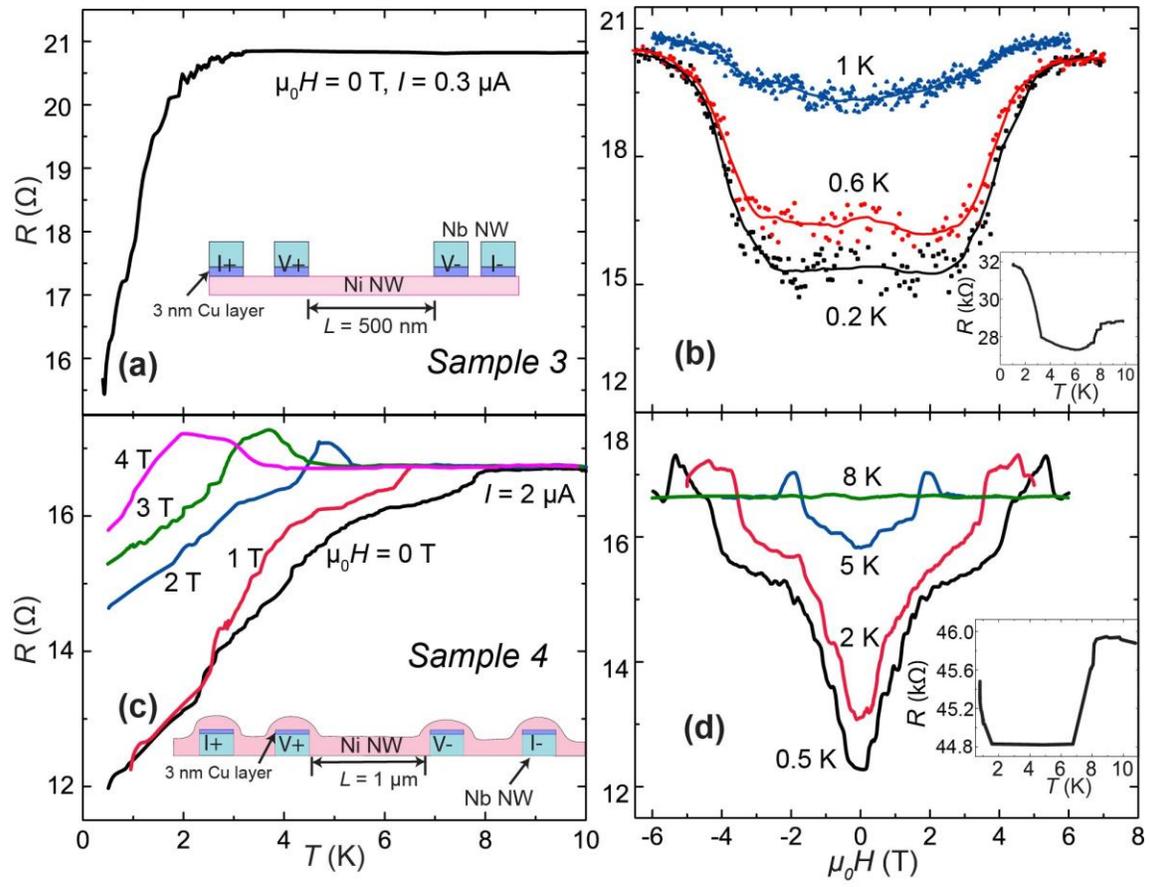

**Figure 2 | *R-T* and *R*-$\mu_0H$ curves of Ni nanowires contacted by Nb electrodes with 3 nm CuO buffer layers.** (a) In *Sample 3*, Ni NW (pink) was thermally evaporated, then it was taken out from vacuum for the second-step lithography and the Cu film (purple) evaporation, it was exposed to air again before placed in a sputtering chamber for Nb (blue) deposition. The resistance *R* starts to drop at *T*~3 K and shows a 25 % decrease at *T* ~ 0.2 K. This suggests a superconducting proximity range of 65 nm in Ni NW. Inset: a schematic drawing of *Sample 3*. (b) *R*-$\mu_0H$ curves of *Sample 3* under different temperatures. The dots are the raw data and the lines are the guides to the eye. Inset: the 2-terminal resistance of *Sample 3*. A resistive upturn shows up below 3 K due to the Mott insulator behavior of the CuO buffer layer. (c) In *Sample 4*, Nb electrodes (blue) along with the Cu (purple) film were sputtered without breaking vacuum, then it was exposed to air and went through the second-step lithography and Ni (pink) NW sputtering. *R-T* curves show a superconductivity proximity range of ~136 nm at zero magnetic field. Inset: a schematic of *Sample 4*. (d) *R*-$\mu_0H$ curves of *Sample 4* at different temperatures. Inset: the 2-terminal resistance of *Sample 4*, a low temperature upturn just like Sample 3 is found. The fabrication details of *Sample 3* and *Sample* 4 are listed in Table 1. Ni NW used here is 400 nm wide and 40 nm thick, while the Nb electrodes are 500 nm wide and 40 nm thick.

**Table 1. Summary of the 8 samples in the superconductivity proximity studies**

| Sample | L (nm) | Fabrication Process | Cu Oxide | Proximity Range along Ni NW (nm) |
|---|---|---|---|---|
| 1 | 500 | Evaporate Ni → Air/lithography → Sputter Nb | No | 2 |
| 2 | 500 | Evaporate Ni → Air/lithography → Evaporate 3 nm Au → Air → Sputter Nb | No | 10 |
| 3 | 500 | Evaporate Ni → Air/lithography → Evaporate 3 nm Cu → Air → Sputter Nb | Yes | 65 |
| 4 | 1000 | Sputter Nb → Sputter 3 nm Cu → Air/lithography → Evaporate Ni | Yes | 136 |
| 5 | 500 | Sputter Ni → Air/lithography → Sputter 3 nm Cu → Sputter Nb | Minimal | 13 |
| 6 | 500 | Sputter Ni → Air/lithography → Sputter 10 nm Cu → Sputter Nb | Minimal | 2 |
| 7 | 500 | Sputter Nb → Air → Sputter 10 nm Cu → Air/lithography → Sputter Ni | Yes | <1 |
| 8 | 1000 | Evaporate Ni → Air → Evaporate 6 nm Au along the entire Ni NW → Air/lithography → Sputter Nb | No | 10 |

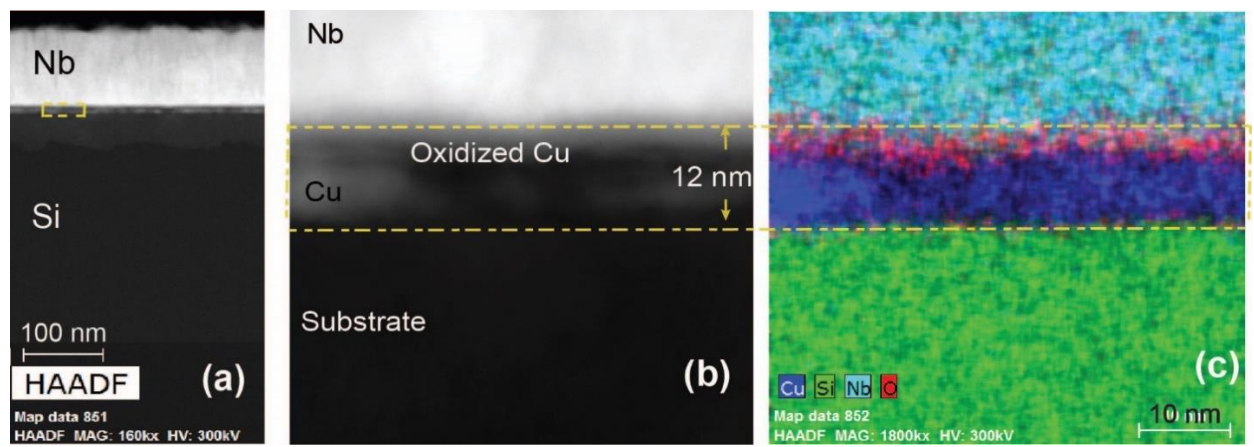

**Figure 3 | Cross-sectional STEM images of the partially oxidized Cu film.** The Nb film was sputtered after the Cu film was exposed to air. (a) Cross-sectional HAADF STEM image. The yellow dashed box spanning over the Nb, Cu and SiO$_2$ regions is highlighted and magnified in (b) and (c). (b) Zoomed-in HAADF STEM image of the magnified region. (c) EDS mappings of the magnified region. It shows an abundance of oxygen, hence CuO, that extends ~ 3 to 4 nm into the 12 nm thick Cu film.

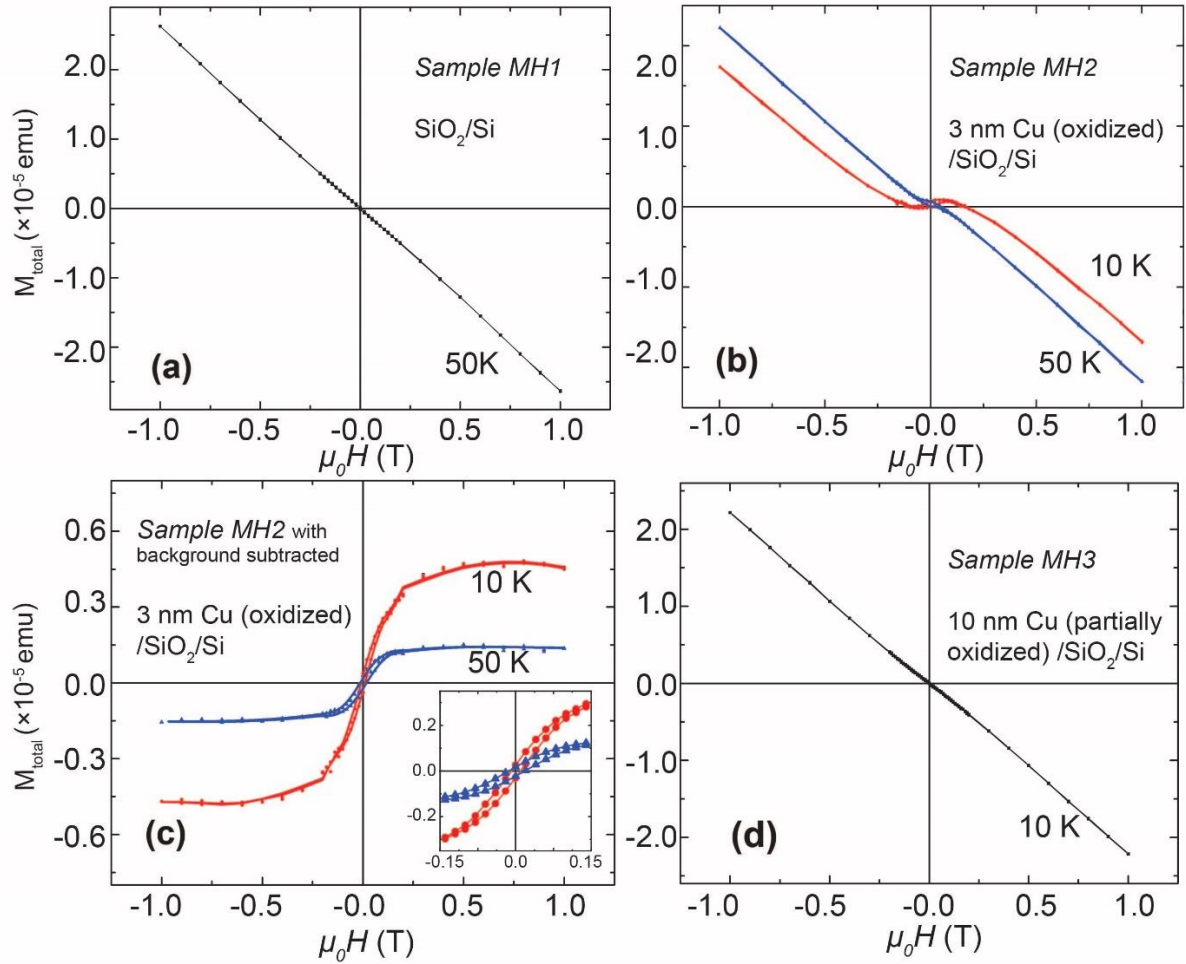

**Figure 4 | *M-H* curves of the SiO$_2$/Si substrate and oxidized Cu films on SiO$_2$/Si substrate**. (a) *M-H* property of a 4 mm × 4 mm SiO$_2$ (500 nm)/Si (500 µm) substrate (*Sample MH1*). The negative linear *M-H* slope demonstrates its diamagnetic property. (b) *M-H* relation of *Sample MH2*. The *M-H* curve shows an inverted sigmoid-shape deviation from diamagnetic behavior for µ$_0$H below 2 kOe. (c) Ferromagnetic response of *Sample MH2* calculated from Fig. 4b via procedures explained in the text. Ferromagnetic hysteresis loops are seen at *T* ~ 10 K and 50 K. The magnetization increases as the temperature is lowered from 50 to 10 K. Inset: Zoomed-in view of the hysteresis loops. (d) *M-H* relation of *Sample MH3*. No ferromagnetic signature is found. The synthesis details of *Sample MH1, Sample MH2,* and *Sample MH3* are listed in Table 2.

**Table 2. Summary of the magnetic property of the substrates and thin evaporated Cu films**

| Thin Film No. | Sample Fabrication | Cu Oxide | Ferromagnetic Response |
|---|---|---|---|
| MH1 | $SiO_2$ (500 nm)/Si (500 μm) | No | No |
| MH2, MH2a, MH2b | $SiO_2$ (500 nm) / Si (500 μm) → Evaporate/Sputter 3 nm Cu → Air | Yes | Yes |
| MH3 | $SiO_2$ (500 nm) / Si (500 μm) → Evaporate 10 nm Cu → Air | Yes, Partially | No |
| MHS1 | $SiO_2$ (500 nm) / Si (500 μm) → Sputter Nb | No | No |
| MHS2 | $SiO_2$ (500 nm) / Si (500 μm) → Sputter Nb → Evaporate/Sputter 3 nm Cu → Air | Yes | Yes |
| MHS3 | $SiO_2$ (500 nm) / Si (500 μm) → Sputter 3 nm Cu → Sputter 3 nm Nb | No | No |

**References**

[1]     R. B. Vandover, A. Delozanne, and M. R. Beasley, *J. Appl. Phys.* **52**, 7327 (1981).

[2]     P. G. Degennes, *Rev. Mod. Phys.* **36**, 225 (1964).

[3]     A. I. Buzdin, *Rev. Mod. Phys.* **77**, 935 (2005).

[4]     Y. N. Chiang, O. G. Shevchenko, and R. N. Kolenov, *Low Temp. Phys.* **33**, 314 (2007).

[5]     R. S. Keizer, S. T. B. Goennenwein, T. M. Klapwijk, G. X. Miao, G. Xiao, and A. Gupta, *Nature* **439**, 825 (2006).

[6]     J. Aumentado and V. Chandrasekhar, *Phys. Rev. B* **64**, 054505 (2001).

[7]     J. W. A. Robinson, J. D. S. Witt, and M. G. Blamire, *Science* **329**, 59 (2010).

[8]     T. S. Khaire, M. A. Khasawneh, W. P. Pratt, Jr., and N. O. Birge, *Phys. Rev. Lett.* **104**, 137002 (2010).

[9]     C. Klose *et al.*, *Phys. Rev. Lett.* **108**, 127002 (2012).

[10]    G. B. Halasz, J. W. A. Robinson, J. F. Annett, and M. G. Blamire, *Phys. Rev. B* **79**, 224505 (2009).

[11]    N. Satchell and N. O. Birge, *Phys. Rev. B* **97**, 214509 (2018).

[12]    J. A. Glick, V. Aguilar, A. B. Gougam, B. M. Niedzielski, E. C. Gingrich, R. Loloee, W. P. Pratt, and N. O. Birge, *Sci. Adv.* **4**, eaat9457 (2018).

[13]    S. Voltan, A. Singh, and J. Aarts, *Phys. Rev. B* **94**, 054503 (2016).

[14]    Y. Wang *et al.*, *arXiv:2012.05898* (2020).

[15]    J. Wang *et al.*, *Nat. Phys.* **6**, 389 (2010).

[16]    F. Bergeret, A. Volkov, and K. Efetov, *Phys. Rev. B* **68**, 064513 (2003).

[17]    S. Takei and V. Galitski, *Phys. Rev. B* **86**, 054521 (2012).

[18]    A. F. Volkov and K. B. Efetov, *Phys. Rev. B* **81**, 144522 (2010).

[19]    M. Kompaniiets, O. V. Dobrovolskiy, C. Neetzel, F. Porrati, J. Brötz, W. Ensinger, and M. Huth, *Appl. Phys. Lett.* **104**, 052603 (2014).

[20]    Y. Pauleau, S. Kukielka, W. Gulbinski, L. Ortega, and S. N. Dub, *J. Phys. D: Appl. Phys.* **39**, 2803 (2006).

[21]    A. K. Singh, U. Kar, M. D. Redell, T.-C. Wu, W.-H. Peng, B. Das, S. Kumar, W.-C. Lee, and W.-L. Lee, *npj Quantum Mater.* **5**, 1 (2020).

[22]    N. Gál, V. Štrbík, Š. Gaži, Š. Chromik, and M. Talacko, *Journal of Superconductivity and Novel Magnetism* **32**, 213 (2019).

[23]   See Supplemental Material at XXXXX for further details regarding more transport results of *Samples 1, 4, 5, 6, 7,* and *8*, and *M-H* curves of thin-film samples.

[24]   H. Yamazaki, N. Shannon, and H. Takagi, *arXiv:1602.05790* (2016).

[25]   J. Wang, C. Shi, M. Tian, Q. Zhang, N. Kumar, J. K. Jain, T. E. Mallouk, and M. H. W. Chan, *Phys. Rev. Lett.* **102**, 247003 (2009).

[26]   O. Skryabina *et al.*, *Sci. Rep.* **9**, 1 (2019).

[27]   K. Shawulienu *et al.*, *Communications Physics* **3**, 116 (2020).

[28]   G. Chen *et al.*, *Phys Rev B* **102**, 115149 (2020).

[29]   Q. Zhang, K. Zhang, D. Xu, G. Yang, H. Huang, F. Nie, C. Liu, and S. Yang, *Prog. Mater. Sci.* **60**, 208 (2014).

[30]   K. Karthik, N. Victor Jaya, M. Kanagaraj, and S. Arumugam, *Solid State Commun.* **151**, 564 (2011).

[31]   D. Gao, J. Zhang, J. Zhu, J. Qi, Z. Zhang, W. Sui, H. Shi, and D. Xue, *Nanoscale Res. Lett.* **5**, 769 (2010).

[32]   D. Gao, G. Yang, J. Li, J. Zhang, J. Zhang, and D. Xue, *J. Phys. Chem. C* **114**, 18347 (2010).

[33]   A. Punnoose, H. Magnone, M. Seehra, and J. Bonevich, *Phys. Rev. B* **64**, 174420 (2001).

[34]   H. Qin, Z. Zhang, X. Liu, Y. Zhang, and J. Hu, *Journal of Magnetism and Magnetic Materials* **322**, 1994 (2010).

[35]   L. Néel, *C.R. Acad. Sci* **252**, 4075 (1961).

[36]   J. Richardson, D. Yiagas, B. Turk, K. Forster, and M. Twigg, *J. Appl. Phys.* **70**, 6977 (1991).